\documentclass[pre,twocolumn,noshowpacs,preprintnumbers,amsmath,amssymb]{revtex4-1}
\usepackage{mathrsfs}
\usepackage[dvipdfmx]{graphicx}
\usepackage{dcolumn}
\usepackage{bm}
\usepackage{color}

\newcommand{\mrA}{\mathrm{A}}
\newcommand{\mrH}{\mathrm{H}}
\newcommand{\mrS}{\mathrm{S}}

\newcommand{\mrC}{\mathrm{C}}

\newcommand{\MSD}{\mathrm{MSD}}
\newcommand{\NGP}{\mathrm{NGP}}

\renewcommand{\vec}{\bm}

\newcommand{\rc}{r_{\rm c}}

\newcommand{\bl}[1]{{\color{blue} #1}}

\newcommand{\SMabb}{SI}

\begin{document}
\title{Loopy L{\'e}vy flights enhance tracer diffusion in active suspensions}

\author	{
			Kiyoshi Kanazawa
		}
		\email{kiyoshi@sk.tsukuba.ac.jp}
\affiliation{
				Faculty of Engineering, Information and Systems, University of Tsukuba, Tennodai, Tsukuba, Ibaraki 305-8577, Japan
			}
\author{Tomohiko G. Sano}
\affiliation{
				Flexible Structures Laboratory, Institute of Mechanical Engineering, 
				\'Ecole polytechnique f\'ed\'erale de Lausanne, Lausanne, CH-1015, Switzerland.
			}
\author{Andrea Cairoli}
\affiliation{Department of Bioengineering, Imperial College London, London SW7 2AZ, UK}
\author{Adrian Baule}
\affiliation{School of Mathematical Sciences, Queen Mary University of London, London E1 4NS, UK }
\date{\today}


\maketitle
{\bf 
	Brownian motion is widely used as a paradigmatic model of diffusion in equilibrium media throughout the physical, chemical, and biological sciences.
	However, many real world systems, particularly biological ones, are intrinsically out-of-equilibrium due to the energy-dissipating active processes underlying their mechanical and dynamical features~\cite{Needleman2017}. 
	The diffusion process followed by a passive tracer in prototypical active media such as suspensions of active colloids or swimming microorganisms \cite{Koch:2011aa} indeed differs significantly from Brownian motion, 
	manifest in a greatly enhanced diffusion coefficient~\cite{Wu2000,Leptos2009,Mino:2011aa,Kurtuldu:2011aa,Mino:2013aa,Jepson:2013aa,Jeanneret2016,Kurihara2017}, 
	non-Gaussian tails of the displacement statistics~\cite{Kurtuldu:2011aa,Jeanneret2016,Kurihara2017}, and crossover phenomena~\cite{Jeanneret2016,Kurihara2017} from non-Gaussian to Gaussian scaling.
	While such characteristic features have been extensively observed in experiments, there is so far no comprehensive theory explaining how they emerge from the microscopic active dynamics. 
	Here we present a theoretical framework of the enhanced tracer diffusion in an active medium from its microscopic dynamics by coarse-graining the 
	hydrodynamic interactions between the tracer and the active particles as a stochastic process. 
	The tracer is shown to follow a non-Markovian coloured Poisson process that accounts quantitatively for all empirical observations.
	The theory predicts in particular a long-lived L{\'e}vy flight regime \cite{Hughes1981} of the tracer motion with a non-monotonic crossover between two different power-law exponents. 
	The duration of this regime can be tuned by the swimmer density, thus suggesting that the optimal foraging strategy of swimming microorganisms might crucially depend on the density in order to exploit the L{\'e}vy flights of nutrients \cite{Bartumeus2002}.
	Our framework not only provides the first validation of the celebrated L{\'e}vy flight model \cite{Hughes1981} from a physical microscopic dynamics, but can also be applied to address important conceptual questions, 
	such as the thermodynamics of active systems~\cite{KanazawaPRL2015}, and practical ones regarding, 
	e.g., the interaction of swimming microorganisms with nutrients and other small particles like degraded plastic~\cite{Goldstein:2015aa} and the design of artificial nanoscale machines~\cite{Bechinger:2016aa}. 
}


	A passive tracer immersed in a fluid at equilibrium moves randomly due to its collisions with the surrounding fluid molecules. 
	Understanding how the observed stochastic process followed by the tracer relates to the statistical mechanics of the surrounding fluid, as accomplished in the seminal works by Einstein, Smoluchowski, and Langevin~\cite{Gardiner1985}, 
	has provided deep insight into the connection between molecular transport and equilibrium thermodynamics
	, which has been widely exploited to describe soft matter and other complex physical systems \cite{Coffey2004}. 
	However, when either artificial self-propelled colloids or biological swimming micro-organisms, such as bacteria like {\it Escherichia coli} or algae like {\it Volvox} and {\it Chlamydomonas reinhardtii}~\cite{Koch:2011aa}, are also suspended, 
	the diffusion of the tracer changes dramatically due to the active stirring of the fluid exerted by the self-propelled particles.
	Indeed, the active diffusion of the tracer experimentally exhibits the following unique features that can no longer be explained as a Brownian motion: 
	(i)~the tracer exhibits dancing loopy trajectories~\cite{Leptos2009,Jepson:2013aa,Jeanneret2016};
	(ii)~its mean square displacement (MSD) exhibits a crossover between superdiffusion with $\sim t^{\alpha}$ ($1<\alpha\leq 2$) 
	for short times and normal diffusion ($\alpha=1$) for long times, 
	where the effective diffusion coefficient $D$ is greatly enhanced compared with that of the equilibrium case, $D_0$, 
	revealing for both three and quasi-two dimensional systems a linear dependence on the 
	density of swimmers $\rho$:	$D=D_0+B\rho$, with	the coefficient $B$ being system dependent~\cite{Wu2000,Leptos2009,Mino:2011aa,Mino:2013aa,Jepson:2013aa,Jeanneret2016};  
	(iii)~the probability density function (PDF) $P_{\Delta t}$ of position displacements in a given time interval $\Delta t$ 
	exhibits strong non-Gaussian features manifest as power-law tails~\cite{Kurtuldu:2011aa,Kurihara2017}; 
	(iv)~$P_{\Delta t}$ eventually reverts to a Gaussian shape for large $\Delta t$ \cite{Jeanneret2016,Kurihara2017}; 
	(v)~the associated non-Gaussian parameter exhibits a scaling regime $\sim \Delta t^{-1}$ for large times \cite{Kurihara2017}.

	Developing a single theory that captures all features (i)--(v) has been a major challenge, 
	due to the multi-particle hydrodynamic, and thus long-range, interactions of the tracer with the swimmers underlying its transport.
	While the loop-like motion (i) results from an individual scattering event of the tracer 
	in the dipolar flow field of a single swimmer~\cite{Dunkel:2010aa,Lin:2011aa,Mino:2013aa},
	and the linear form of $D$ (ii) has been explained phenomenologically based on the active flux of the swimmers, 
	which is defined as the product of their number density and characteristic swimming speed~\cite{Mino:2011aa,Mino:2013aa,Jepson:2013aa,Lin:2011aa,Morozov:2014aa},
	the statistical observations (iii)--(v) could so far not be explained consistently. 
	The power-law tails in $P_{\Delta t}$ and their convergence to Gaussian scaling for long observation times, 
	which is expected based on the central limit theorem (CLT) arguments~\cite{Pushkin:2013aa,Thiffeault2015},
	have been reproduced in \cite{Zaid2011,Zaid2016,Kurihara2017} assuming a static force distribution 
	akin to the Holtsmark theory of gravitating particles (see {\SMabb} for a review)~\cite{Holtsmark1919}. 
	However, this approach neglects any dynamics of the swimmers and 
	is thus not sufficient to capture the enhanced diffusion 
	observed in experiments~\cite{Wu2000,Leptos2009,Mino:2011aa,Kurtuldu:2011aa,Mino:2013aa,Jepson:2013aa,Jeanneret2016,Kurihara2017}.
	Here, we present a derivation of the stochastic process 
	underlying the enhanced diffusion of the tracer from microscopic dynamics that is valid at all timescales. 
	The resulting process captures all characteristic features (i)--(v) and is in excellent quantitative agreement with simulation data.

\begin{figure*}[p]
		\centering
		\includegraphics[width=180mm]{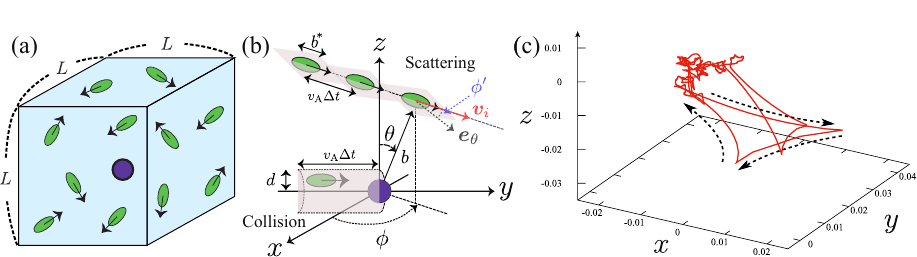}
		\caption{
			(a)~Schematic of the microscopic model: 
				active particles (green ellipsoids) and a passive tracer (filled violet sphere) are suspended in a cubic box of linear length $L$. 
				The direction of the active swimmers $\bm n_i \equiv (\sin{\theta_i}\cos{\phi_i},\sin{\theta_i}\sin{\phi_i},\cos{\theta_i})$ is randomized upon hitting the box boundary. 
				We finally take the large system-size limit keeping the number density of active particles $\rho\equiv N/L^3$ constant (see {\SMabb} for protocol details).  
				The bead is dragged by the hydrodynamic flow fields generated by the swimming of active particles.
			(b)~Exemplary two-body scattering event with impact parameter $b>d$ and injection angles $\theta$, $\phi$, and $\phi'$, 
				unit vector $\bm{e}_\theta$, and collision event with $b\leq d$. 
				The force shape function during scattering is thus characterized by the parameter set $\bm{b}\equiv (b,\theta,\phi,\phi')$. 
				An active particle travels the distance $v_{\mrA} \Delta t$ in a time interval $\Delta t$. 
				This distance is equal to the characteristic lengthscale $b^*$ at the time $\tau_{\mrH}$ such that $v_{\mrA}\tau_{\mrH}=b^* \Longleftrightarrow \tau_{\mrH}\equiv b^*/v_{\mrA}$. 
				Since $b^*$ is related to the typical length scale of the active swimmers, for $\Delta t\ll \tau_{\mrH}$
				their motion can be effectively neglected (see \SMabb). 
				Conversely, for a given set of injection angles active particles can collide against the tracer in the time interval $\Delta t$ if they are contained 
				in a cylinder with cross section area $\pi d^{2}$ and linear length $v_{\mrA} \Delta t$ surrounding the passive particle.	
				The mean free time of the tracer $\tau_{\mrC}$ is then estimated as 
				$\rho \pi d^{2} v_{\mrA}\tau_{\mrC}=1 \Longleftrightarrow \tau_{\mrC}\equiv 1/\rho \pi d^{2} v_{\mrA}$. 
				For $\Delta t \ll \tau_{\mrC}$ such collision events do not contribute to the tracer dynamics. 
			(c)~Typical trajectory of the passive tracer observed in simulations, exhibiting its dancing L\'evy motion in the case of pushers ($p<0$ and $d=b^*$). 
				Once scattered by an active swimmer, the tracer exhibits a triangular-shaped but non-closed trajectory 
				in the plane specified by the injection angles of the active swimmer $\theta$, $\phi$, and $\phi^{\prime}$ (dashed lines). 
				We choose $b^*$ as unit for lengths.
		}
		\label{fig:setup}
		
		\par
		\vspace{30pt} 
		
		\centering
		\includegraphics[width=180mm]{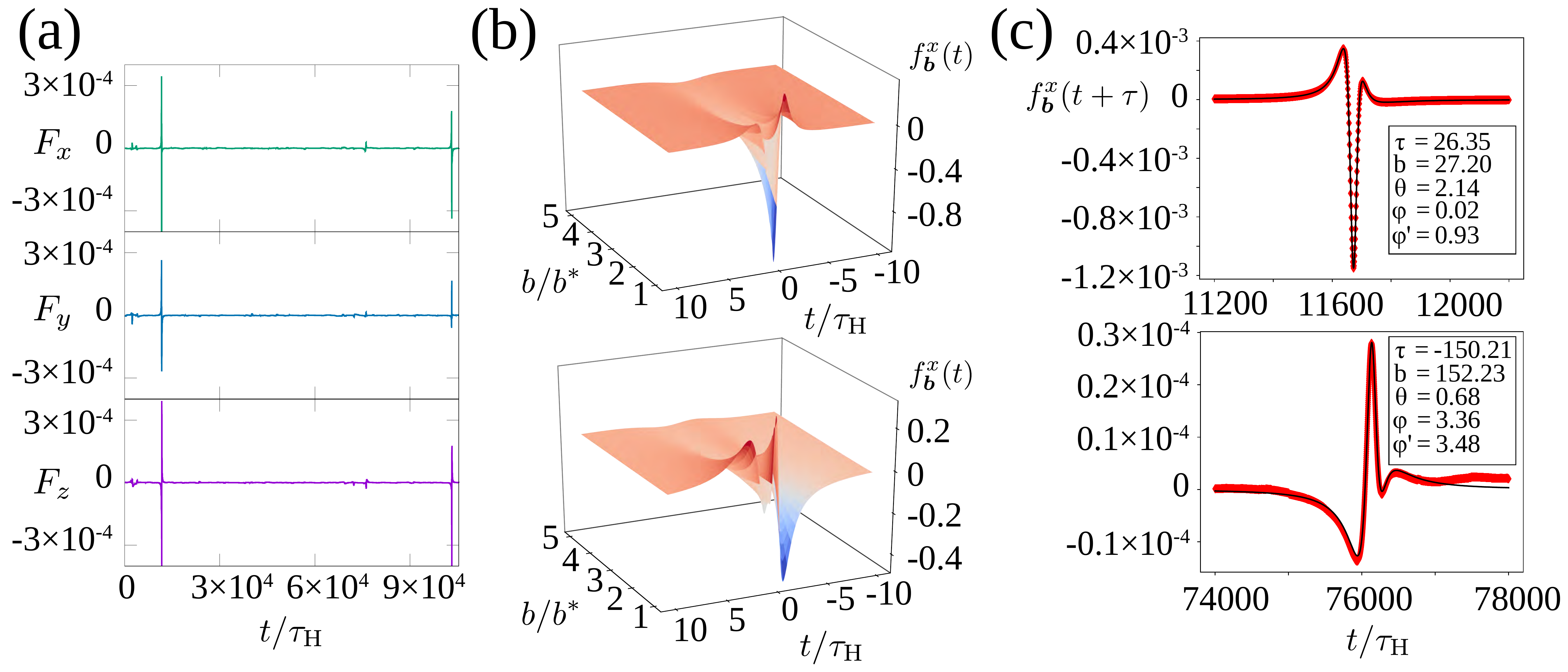}
		\caption{
				We choose $b^*$, $\tau_{\mrH}$ as units for length and time respectively.
			(a)~Typical time series of force exerted on the tracer by the active swimmers in dilute conditions under the parameters $p<0$ (pushers) and $d=b^*$. 
			(b)~3D plots of the force shape function for the stresslet hydrodynamic force in Eq.~\eqref{force}, 
				obtained by perturbatively solving the two-body scattering problem ({\SMabb}).  
				We set $\Gamma v_A=1$, and $\{\theta,\phi,\phi^{\prime}\}=\{\pi/2,0,\pi/2\}$ (top panel) or $\{\theta,\phi,\phi^{\prime}\}=\{0,0,0\}$ (bottom panel).
				The general formula is obtained by linearly combining these two solutions weighted by coefficients that depend only on the angular variables. 
			(c)~Exemplary fits (black solid line) of the force shape function $f_{\vec{b}}(t)$ to simulation data of force exerted on the tracer in the direction $\vec{e}_x$ 
				during different scattering events arbitrarily extracted from force time-series. Fit parameters are the set $\vec{b}$ and the scattering time $\tau$.
				The fit is obtained by using non-linear least squares.  			
				The agreement is excellent. 
		}
		\label{fig:fsf}
	\end{figure*}

	We consider a three-dimensional system composed of $m$ active particles and a passive tracer suspended in a viscous fluid in a cubic box of linear length $L$ (Fig.~{\ref{fig:setup}}a). 
	The active particles (swimmers) are assumed self-propelled and their motion is driven by unidirectional forces with constant amplitude~\cite{Lauga2009}. 
	In general, the dynamics of such multi-particle systems where interactions are mediated by the fluid environment in the form of hydrodynamic forces are complex and analytically intractable.
	However, suspensions of micro-organisms often considered in experiments (see above) are generally characterized by (a) low Reynolds number swimming and (b) a low density of swimmers (dilute conditions, see below).
	In particular the dilute condition (b) allows us to neglect the mutual hydrodynamic interactions of the swimmers \cite{Drescher2011}, thus leading to the overdamped equations of motion:
	\begin{align}
		\frac{d\bm x_i}{dt} &= v_{\mrA}\bm n_i , &
		\Gamma \frac{d\bm X}{dt}   &= \sum_{i=1}^m \bm F(\bm x_i-\bm X,\bm n_i), 
		\label{model}		
	\end{align}
	where $\bm{F}$ is the force on the tracer generated by a single swimmer and $\Gamma$ is a viscous coefficient for the passive particle. 
	In Eq.~\eqref{model}, $\bm x_i(t)$ and $\bm X(t)$ denote the positions of the $i$-th active particle and the passive particle, respectively, 
	and $v_{\mrA}$ is the constant amplitude of the active self-propulsion force with unit vector $\bm n_i$ specifying the swimming direction.

	The low Reynolds number condition (a) further yields a closed form expression for $\bm{F}$ as the solution of the Stokes equation. 
	For force- and torque-free swimmers, the leading-order term of this solution in a far-field expansion is a dipole force \cite{Lauga2009}. 
	Therefore, $\bm F$ is specified without loss of generality as the stresslet
	\begin{equation}
		\bm F(\vec{r}_i, \bm n_i) \approx \dfrac{p}{r_i^2}\left[3\dfrac{(\bm n_i \cdot \vec{r}_i)^2}{r_i^2}-1 \right]\dfrac{\vec{r}_i}{r_i}
		\label{force}
	\end{equation}
	for $r_i>d$, with the difference vector $\vec{r}_i=\bm x_i-\bm X$, the dipole strength $p$ 	and the system specific cut-off $d$.  
	The dipole strength parameter $p$ specifies the universal features of the far-flow hydrodynamic field \cite{Lauga2009}: 
	$p<0$ denotes \textit{pusher} swimmers whose flow lines are oriented outward along the direction of its velocity vector and inward laterally (e.g., \textit{E. coli}~\cite{Lauga2009}); 
	$p>0$ denotes instead \textit{puller} swimmers whose flow lines are oriented in the opposite directions (e.g., \textit{Chlamydomonas}~\cite{Lauga2009}).	
	The length parameter $d$ separates the far-flow field from any near-flow field hydrodynamic contributions and hard-core interactions, 
	and thus determines when the approximation~\eqref{force} is valid. 
	The model has one further length parameter  
	$b^* \equiv \sqrt{|p|/\Gamma v_{\mrA}}$, which can be related to the typical length scale of the swimmer~\cite{Lauga2009}.
	All these parameters can be determined experimentally: for \textit{E. coli} $d\simeq 6\, \mu\text{m}$ and $b^*\simeq 2\, \mu\text{m}$  \cite{Mino:2011aa,Drescher2011};
	for \textit{Chlamydomonas} $d\simeq 35\, \mu\text{m}$ and $b^*\simeq 8\, \mu\text{m}$  \cite{Drescher2010,Zaid2011,Zaid2016,Kurihara2017},
	satisfying $d\geq b^*$.
	
	Conversely, for $r_i\leq d$ the interaction force $\bm F$ is not universal but system-specific. 
	Nevertheless, all swimmer-tracer interactions in this regime can be accurately captured by using arguments based on the CLT (see below), 
	which do not require a detailed form of $\bm{F}$.
	Therefore, in our present implementation we set for simplicity $\bm{F}=0$ in this regime, 
	but we remark that our qualitative results are independent of this specific choice  
	and that this approximation has also been validated experimentally~\cite{Kurihara2017}. 

\begin{figure*}[!htb]
		\centering
		\includegraphics[width=180mm]{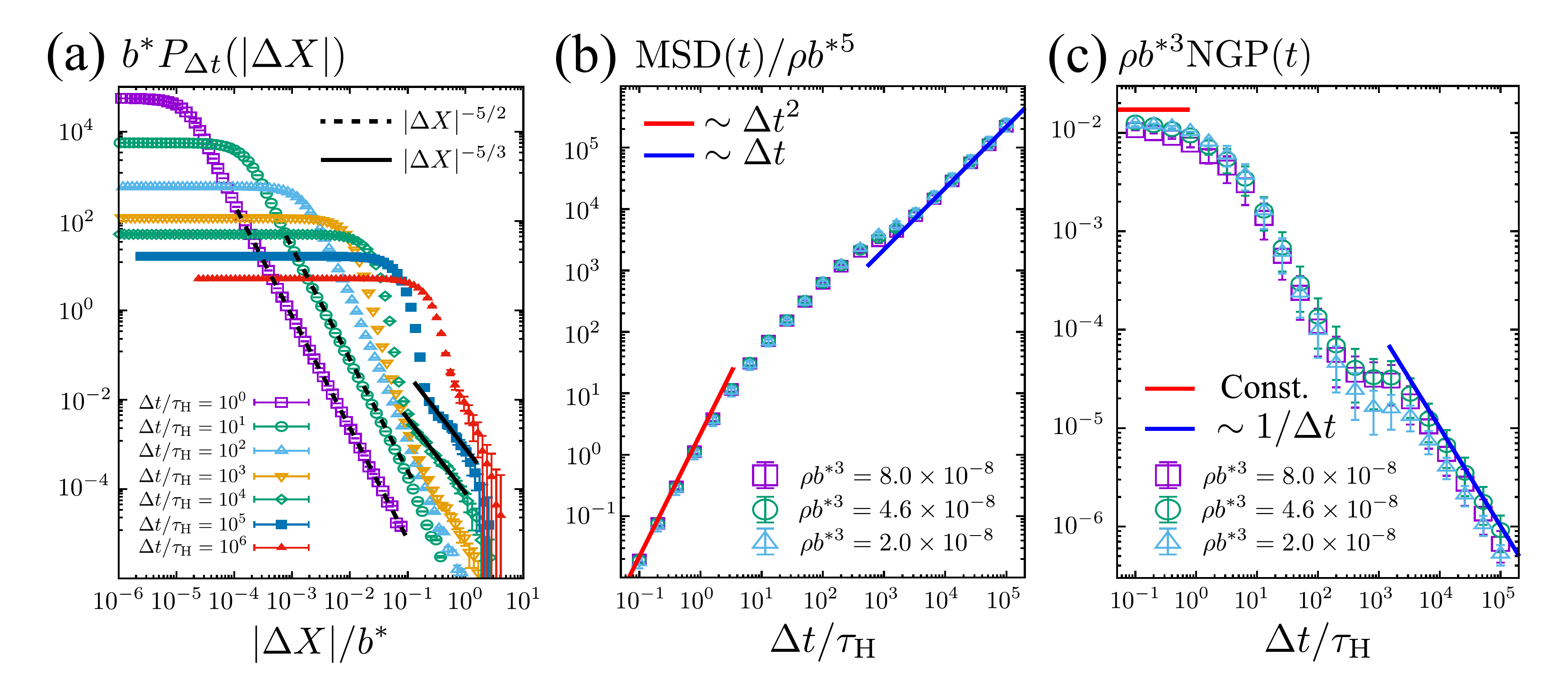}
		\caption{
			(a)~The numerical distribution $P_{\Delta t}(|\Delta X|)$ for the displacement of the tracer $\Delta X$, showing power-law tails. 
				The tails particularly exhibit a non-monotonic behaviour, becoming fatter from the Holtsmark regime ($\Delta t\ll \tau_{\mrH}$), 
				where $P_{\Delta t}\sim |\Delta X|^{-\bl{\alpha_{\mathrm{H}}}}$, to the scattering regime ($\tau_{\mrH}\ll \Delta t\ll \tau_{\mrC}$), 
				where instead $P_{\Delta t}\sim |\Delta X|^{-\alpha_{\mathrm{S}}}$. 
				This striking behaviour of the enhanced diffusion process is predicted by our theory (see Eq.~{(\ref{displacement})}) 
				and should be observable experimentally.
			(b)~The mean-squared displacement is plotted for various swimmer densities $\rho$, 
				and shown to collapse onto a single master curve upon rescaling by $1/\rho b^{* 5}$. 
				One observes a clear crossover between ballistic and normal diffusive motion \cite{Wu2000,Leptos2009,Mino:2013aa,Kurtuldu:2011aa,Kurihara2017}.
			(c)~The non-Gaussian parameter rescaled by $\rho b^{*3}$ is plotted for various densities, 
				thus showing the universal power-law decay $\Delta t^{-1}$ 
				consistently with the experimental observations in \cite{Kurihara2017}.
		}
		\label{fig:results}
	\end{figure*}

	Numerical simulations of Eqs.~(\ref{model},\ref{force}) in dilute conditions reproduce all features~(i)--(v)
	(see Fig.~\ref{fig:setup}c and Fig.~\ref{fig:results}). 
	Crucially, an inspection of time-series trajectories under such dilute conditions shows that the dynamics of the tracer can be resolved as a sequence of individual scattering events, 
	where only two-body tracer-swimmer interactions are relevant (Fig.~\ref{fig:fsf}a. See also \SMabb). 
	In the kinetic theory of gases the dilute condition is ensured by requiring $\rho \rc^3\ll 1$, where $\rho$ is the number density and $r_{\rm c}$ the range of interparticle interactions. 
	In our system, however, the hydrodynamic force is long-range such that an interaction range cannot be well defined. Nevertheless, we can identify $\rc$ with the maximum geometrical length parameter 
	as $\rc \equiv \max\{d, b^*\} = d$ and define dilute conditions accordingly. 
	The motivation behind this definition of $\rc$ is two-fold. 
	Firstly, with this definition the condition $\rho d^{3}\ll 1$ is indeed realized in experiments that exhibit the features (i)--(v) 	
	(e.g., $\rho d^3 \sim 0.15$ in \cite{Kurihara2017} and $\rho d^3 \sim 0.1$ in \cite{Mino:2011aa}).
	Secondly, this parameter regime allows for a self-consistent description of the tracer dynamics, in which the dipole interaction governs the displacement statistics on short and intermediate time scales, 
	while the statistics for longer times reverts to a Gaussian form due to the CLT.
	To this end, we note that in a dilute system, at every instant in time, swimmers on average have $b\gg d$,  
	where $b$ is introduced as the impact parameter of a binary swimmer-tracer interaction (see Fig.~\ref{fig:setup}b). 
	The tracer statistics will then be governed by three distinct dynamical regimes: 
	(1.) For short times $\Delta t\ll \tau_{\mrH}$, the tracer experiences the long-range forces of effectively static swimmers as in the Holtzmark theory (``Holtzmark regime"). 
	Accordingly, $\tau_{\mrH}\equiv b^*/v_{\mrA}$. 
	(2.) For times $\tau_{\mrH}\ll \Delta t \ll \tau_{\mrC}$, the tracer is scattered by the moving swimmers in a sequence of binary interactions that we call ``scatterings" (``scattering regime"). 
	Since the system is dilute, the swimmer-tracer interaction is governed by the far-flow field as in Eq.~\eqref{force} in this regime. 
	The time scale $\tau_{\mrC}\equiv 1/\rho v_{\mrA} \pi d^{2}$ (see Fig.~\ref{fig:setup}b) thus estimates the time necessary for a swimmer to arrive close enough to the tracer to interact via hard-core and near-field hydrodynamic interactions. 
	(3.) For $\Delta t\gg \tau_{\mrC}$, these interaction events thus represent ``collisions" and the tracer is displaced by an accumulation of such collisions with the swimmers such that the CLT applies (``CLT regime").
	
	Remarkably, these three regimes are captured by a coarse-grained description of the tracer dynamics in terms of the Langevin equation
	\begin{align}
		\Gamma \frac{d \bm X}{dt} = \bm F(t), \>\>\>\>
		\bm F(t) \equiv \sum_{i=1}^{N(t)} \bm f_{\bm {b}_i}(t-\tau_i),
		\label{eq:colP}
	\end{align} 
	where $N(t)$ counts the number of scattering events up to a fixed time $t$ and $\bm f_{\bm{b}}(t)$ is the force shape function (FSF) describing the force exerted on the tracer during each scattering.
	The transition from the fully deterministic dynamics of Eqs.~(\ref{model}, \ref{force}) to a stochastic description by Eq.~\eqref{eq:colP} is realised by assuming $N(t)$ to be a Poisson process with intensity $\lambda(\bm{b})$.
	The FSF $\bm f_{\bm{b}}(t)$ is characterized by the set of impact parameter and injection angles (denoted by $\bm{b}$ in Fig.~\ref{fig:fsf}b) and is obtained in analytical form by solving the binary swimmer--tracer scattering problem.
	Remarkably, for $b\gg b^*$ (valid on average in dilute conditions and in the Holtzmark and scattering regimes, see above)
	an analytical approximation of the FSF can be obtained using a Picard iteration up to 2nd order and subsequent Taylor expansion (see {\SMabb}), which is in excellent agreement with numerics (Fig.~\ref{fig:fsf}c). 
	The FSF is centred at the scattering time point $\tau$, which is set by the condition $\bm n \cdot \bm r=0$.
	The intensity can be shown to satisfy $\lambda(\bm{b})\propto b$ directly from its microscopic dynamics (see {\SMabb}), 
	and the total intensity diverges in the large system size limit as $\lim_{L\to \infty}\int d\bm{b} \lambda(\bm{b}) = \infty$.
	Consequently, ${\bm F}$ is a coloured Poisson noise with infinite intensity, also known as generalized Campbell's process \cite{Campbell1909}. 
	This infinite intensity is a typical singular character of a L{\'e}vy process, and is a physical consequence of the long-range hydrodynamic interactions, 
	which cause an infinite number of small scatterings at possibly infinite impact parameter.
				
	We now calculate the statistics resulting from Eq.~\eqref{eq:colP} for the tracer displacement in the $x$-direction 
	$\Delta X \equiv [\bm{X}(t+\Delta t)-\bm{X}(t)] \cdot \bm{e}_x$   
	in agreement with the experimental protocols typically used \cite{Leptos2009,Kurtuldu:2011aa,Jeanneret2016,Kurihara2017}.
	Using functional techniques (see {\SMabb}), we derive the displacement PDF $P_{\Delta t}(|\Delta X|)$ and obtain its scaling behaviours in the regimes (1.)--(3.) as
	\begin{equation}
		P_{\Delta t}(|\Delta X|) \propto 
		\begin{cases}
			|\Delta X|^{-\alpha_{\mrH}} & (\Delta t \ll \tau_{\mrH}) \\
			|\Delta X|^{-\alpha_{\mrS}} & (\tau_{\mrH} \ll \Delta t \ll \tau_{\mrC}) \\
			e^{-\Delta X^2/2\sigma^2} & (\tau_{\mrC} \ll \Delta t)
		\end{cases}
		\label{displacement}
	\end{equation}
	with power-law exponents $\alpha_{\mrH}=5/2$, $\alpha_{\mrS}=5/3$ and positive constant $\sigma^2$ that depends on $d$ (see {\SMabb}). 
	For finite $d$ a truncation appears in the power-law tails, which is realistic because any physical system must accommodate finite cutoffs \cite{Mantegna1994}.

	The Holtsmark regime (1.) yields the same scaling behaviour as found in other purely static approaches $|\Delta X|^{-\alpha_{\mrH}}$~\cite{Zaid2011,Zaid2016,Kurihara2017}. 
	For $\Delta t \gtrsim \tau_{\mrH}$, these theories are not applicable because the rearrangements of the active swimmers are no longer negligible. 
	In fact, the force induced during binary scatterings has to be fully taken into account, 
	and the stochastic process $X$ is then non-Markovian (see {\SMabb}). 
	In the scattering regime (2.), the description nevertheless becomes effectively Markovian 
	as the FSF can be approximated effectively as a Dirac $\delta$-function (see {\SMabb}). 
	In this regime the coloured Poisson model~{(\ref{eq:colP})} is equivalent to a compound Poisson process with jump-length distribution prescribed 
	as a truncated power-law with scaling behaviour $|\Delta X|^{-\alpha_{\mrS}}$. 
	Our results thus validate the well known L\'evy flight model \cite{Hughes1981} 
	as an approximate description of the tracer dynamics at this timescale. 
	A related jump-diffusion process underlying the enhanced diffusion has previously been proposed in \cite{Jeanneret2016} on purely phenomenological grounds. 
	In the collision regime (3.), collisions become dominant over scatterings. 
	Since collisional impact has finite cutoff, accumulation of a sufficient number of collisions leads to the Gaussian tail as a consequence of the central limit theorem.
	We note that the detailed form of $\bm{F}$ for $r_i\leq d$ is renormalized into the variance $\sigma^2$, though it is irrelevant to the power-law tail (e.g., $\alpha_{\mrH}$ and $\alpha_{\mrS}$).
	Overall, the striking non-monotonic behaviour of the scaling exponents of the tracer displacement statistics as predicted by Eq.~\eqref{displacement} is in excellent agreement with simulation results (Fig.~{\ref{fig:results}}a).

	Using the exact expression for $P_{\Delta t}$ ({\SMabb}), the tracer MSD for different concentrations
	of active swimmers can be shown to collapse onto the same universal curve upon rescaling by $1/\rho b^{*5}$, 
	which reveals a crossover from ballistic motion $\sim (\Delta t/\tau_{\mrH})^2$ 
	for short timescales to normal diffusion $\sim \Delta t/\tau_{\mrH}$ for longer ones (see Fig.~\ref{fig:results}b). 
	This yields in the ballistic regime: $\MSD(\Delta t)\propto \rho v_{\mrA}^2 \Delta t^2$.
	Likewise, in the diffusive regime we find: $\MSD(\Delta t)\propto \rho v_{\mrA}\Delta t$. 
	Eq.~\eqref{eq:colP} thus predicts the enhancement of the tracer diffusive motion at all time scales. 
	While in the diffusive regime it captures the linear dependence of the diffusion coefficient $D$ on the active flux $\rho v_{\mrA}$ 
	that has been validated experimentally~\cite{Wu2000,Leptos2009,Mino:2011aa,Mino:2013aa,Jepson:2013aa,Jeanneret2016}, 
	in the ballistic regime it predicts a dependence of the $\MSD$ on $\rho v_{\mrA}^2$.

	Likewise, we can also calculate the non-Gaussian parameter for the tracer displacement statistics 
	$\NGP(\Delta t)\equiv \langle\Delta X^4 \rangle/3\langle \Delta X^2 \rangle^2-1$. 
	Upon rescaling by $\rho b^{*3}$, simulation data for different concentrations
	of active swimmers indeed collapse onto the same curve, 
	that scales as $(\Delta t/\tau_{\mrH})^{-1}$ for long observation times (see Fig.~\ref{fig:results}c). 
	This scaling prediction	is confirmed by the model that yields    
	$\NGP(\Delta t)\propto (\rho v_{\mrA})^{-1}\Delta t^{-1}$ (see \SMabb).
	
	The tracer is also subjected to thermal fluctuations 
	that can be included in the model~\eqref{eq:colP} as Brownian motion 
	(with variance $2 k_{\mathrm{B}}\Gamma T$, the Boltzmann constant $k_{\mathrm{B}}$ and $T$ the temperature of the bath without swimmers). 
	While the scaling predictions~\eqref{displacement} for the tracer displacement distribution and that for the $\NGP$ are preserved, 
	thermal noise breaks the data collapse for both the $\MSD$ and the $\NGP$ 
	and shifts the $\MSD$ by a term $2 D_0 \Delta t$ 
	with $D_0\equiv k_{\mathrm{B}} T/\Gamma$ (see {\SMabb} for details).
	
	Our approach can be extended straightforwardly to any dimensional systems. Further, more general hydrodynamic force fields can be readily incorporated than Eq.~\eqref{force}, 
	such as higher-order terms in the far-field expansion (e.g., the quadrupole).
	Let us assume the scaling relations for the far-flow hydrodynamic interaction force $F\propto 1/r_i^{n_{\mrH}}$ (or equivalently, $f_{\bm{b}}\propto 1/b^{n_{\mrH}}$) 
	and for the net tracer displacement during a scattering event $\int_{-\infty}^{\infty}f_{\bm{b}}(t)dt \propto 1/b^{n_{\mrS}}$ (see \SMabb).
	As a general recipe, our theory predicts power-law crossover with $\alpha_{\mrH}=D/n_{\mrH}$ and $\alpha_{\mrS}=(D-1)/n_{\mrS}$, considering that $\lambda(\bm b)|J_{\bm b}|\propto b^{D-2}$ for $D$-dimensional active flows.
	This suggests that, while $\alpha_{\mrH}$ is universally determined by the power-law exponent of the hydrodynamic interaction force, 
	$\alpha_{\mrS}$ depends crucially on its detailed form (i.e., the existence of loops).

	The enhanced diffusion process as determined by (i)--(v) is a linear-in-time diffusion process with non-Gaussian displacement statistics. 
	Such processes represent a ubiquitous feature of non-equilibrium diffusion phenomena \cite{Wang:2012aa}, 
	but yet they pose considerable challenges already for purely phenomenological modelling approaches \cite{Chechkin:2017aa}.  
	The fact that the stochastic process Eq.~\eqref{eq:colP} captures all its characteristic features is thus remarkable, 
	even more so since it is derived through an exact coarse-graining of microscopic dynamics.
	In contrast to many models that solve this ``Brownian yet non-Gaussian" diffusion conundrum, 
	Eq.~\eqref{eq:colP} is easily interpretable physically due to the well-defined force pulse nature. 
	Models of this type might thus be applicable much more generally.	
	A striking prediction of our theory is the intermediate fatter tail of the tracer displacement distribution
	in the scattering regime compared with the Holtsmark regime, 
	which is counter-intuitive since the tails are typically expected to become thinner monotonically for longer times due to the CLT. 
	Such signature should be experimentally detectable. 
	Moreover, our results make predictions 
	on the possible foraging behaviour of real swimming micro-organisms like \textit{Chlamydomonas}.
	On the one hand, $\tau_{\mrC}$ diverges for $\rho\to 0$, which implies 
	that the power-law tail persists for longer timescales the more dilute the system is. 
	On the other hand, L\'evy flights have been shown to increase encounter probabilities 
	in the stochastic search for sparsely and randomly distributed revisitable targets~\cite{Viswanathan1999,Humphries:2012aa}. 
	These results thus suggest that for swimming micro-organisms an optimal foraging strategy should significantly depend on the density. 
	For large densities the displacements of nutrients are primarily Gaussian (i.e. spatially localized), 
	which make an active search like intermittent L{\'e}vy-Brownian strategies \cite{Benichou2011} more efficient for the forager.  
	On the contrary, at low densities such displacements are power-law distributed, 
	such that it might be advantageous for the forager to simply wait for a nutrient to come close dragged by the other swimmers \cite{Bartumeus2002}.
	Finally, we remark that superimposing additional force fields might lead to novel mechanisms to control and exploit enhanced diffusion in artificial devices.


\begin{acknowledgments}
	We appreciate D.~Mizuno, H.~Takayasu, M. Takayasu, H.~Hayakawa, and F.~van Wijland for fruitful discussions. 
	This work was supported by Grant-in-Aid for JSPS Fellows (Grant No.~16J05315), 
	JSPS KAKENHI (Grant Nos.~16K16016 and 18K13519), 
	the Research Fellowship granted by the Royal Commission for the Exhibition of 1851, 
	and Atoms program granted by the Yukawa Institute for Theoretical Physics. 
	The numerical calculations were carried out on XC40 at Yukawa Institute for Theoretical Physics in Kyoto University.
\end{acknowledgments}

\section*{Author contributions}
	KK conceived the original idea. KK, TGS, AC and AB designed research. KK and AC performed analytical calculations. TGS performed numerical simulations. KK, AC, and AB wrote the paper.

\section*{Competing interests}
	The authors declare no competing financial interests. 

\section*{Data availability}
	We will provide all the numerical codes and plot files upon requests. 




\end{document}